\def\ra{\rangle}
\def\la{\langle}
\def\be{\begin{equation}}
\def\ee{\end{equation}}
\def\ba{\begin{array}}
\def\ea{\end{array}}

\documentclass[aps,pre,preprint,showpacs,amsmath,amssymb,amsfonts,preprintnumbers]
{revtex4}
\usepackage{epsfig,graphicx}
\usepackage{amsmath}

\begin{document}

\baselineskip=18pt \setcounter{page}{1} \centerline{\large\bf Bounds on Multipartite Concurrence and Tangle}
\medskip \vspace{4ex}
\begin{center}
Jing Wang$^{1,2}$, Ming Li$^{1}$, Hongfang Li$^{1}$, Shao-Ming
Fei$^{2,3}$ and Xianqing Li-Jost$^{3}$

\vspace{2ex}

\begin{minipage}{5in}

{\small $~^{1}$ College of the Science, China University of
Petroleum, 266580 Qingdao}

{\small $~^{2}$ Department of Mathematics, Capital Normal
University, 100037 Beijing}

{\small $~^{3}$ Max-Planck-Institute for Mathematics in the
Sciences, 04103 Leipzig}
\end{minipage}
\end{center}

\begin{center}
\begin{minipage}{5in}
\vspace{1ex} \centerline{\large Abstract}
\vspace{1ex}
We present an analytical lower
bound of multipartite concurrence based on the generalized
Bloch representations of density matrices. It is shown that the
lower bound can be used as an effective entanglement witness of genuine multipartite
entanglement. Tight lower and upper bounds for multipartite tangles are also derived.
Since the lower bounds depend on just part of the correlation tensors, the result is experimentally feasible.

\vfill
\smallskip
\end{minipage}\end{center}
\bigskip

\section{Introduction}\label{sec1}

Quantum entanglement, as the remarkable nonlocal feature of quantum
mechanics, is recognized as a valuable resource in the rapidly
expanding field of quantum information science, with various
applications \cite{nielsen, di} such as quantum computation ,
quantum teleportation, dense coding, quantum cryptographic schemes,
quantum radar, entanglement swapping and remote states preparation.

The mixed state $\rho\in\mathcal{H}=\mathcal{H}_{1}\otimes\cdots\otimes\mathcal{H}_{N}$ is said to be a fully separable state, if there exist
$|\phi_{i}^{j}\rangle\in\mathcal{H}_{j},j=1,\ldots,N,~~p_{i}>0,~~\sum_{i}p_{i}=1$, such that
\be
\rho=\sum_{i}p_{i}|\phi_{i}^{1}\rangle\langle\phi_{i}^{1}|\otimes\cdots\otimes|\phi_{i}^{N}\rangle\langle\phi_{i}^{N}|,
\ee
otherwise $\rho$ is said to be an entangled state.
States
that are not biseparable with respect to any partitions are said
to be genuinely multipartite entangled. Genuinely multipartite entanglement is a kind of
important type of entanglement, which offers significant
advantage in quantum information processing tasks \cite{mule1}.
In particular, it is the basic ingredient in
measurement-based quantum computation \cite{mule2}, and is
beneficial in various quantum communication protocols \cite{mule3},
including secret sharing \cite{mule4} (cf.\cite{mule5}).
Despite its importance, characterization and detection of this kind
of resource turn out to be rather hard and only a few results have been
proposed \cite{12, vicente3, 14, 15}.

Quantifying quantum entanglement is a basic and longer standing problem in quantum
information theory. A measure of quantum entanglement can
be used to detect and classify entanglement of quantum states. In this paper,
we use the multipartite concurrence \cite{multicon} to investigate the multipartite
entanglement.  Let ${\mathcal {H}}_{i}$, $i=1,2,\cdots, N$, be $d$-dimensional vector spaces.
The concurrence of an $N$-partite pure state $|\psi\ra\in {\mathcal
{H}}_{1}\otimes{\mathcal {H}}_{2}\otimes\cdots\otimes{\mathcal{H}}_{N}$ is defined by
\begin{eqnarray}\label{xxx}
\mathcal{C}_{N}(|\psi\ra\la\psi|)=2^{1-\frac{N}{2}}\sqrt{(2^{N}-2)-\sum_{\alpha}Tr\{\rho_{\alpha}^{2}\}},
\end{eqnarray}
where $\alpha$ labels all the different reduced density matrices. If we
list all the $2^N-2$ reduced matrices in the following way:
$\{\rho_1,
\rho_2,\cdots,\rho_N,\rho_{12},\rho_{13},\cdots,\rho_{1N},\rho_{23},\cdots,\rho_{12\cdots
N-1},\cdots,\rho_{23\cdots N}\}$, (\ref{xxx}) can be reexpressed as
\begin{eqnarray}\label{xxxx}
\mathcal{C}_{N}(|\psi\ra\la\psi|)=2^{1-\frac{N}{2}}\sqrt{(2^{N}-2)-2\sum_{k=1}^{2^{N-1}-1}Tr\{\rho_{k}^{2}\}}.
\end{eqnarray}
For a mixed multipartite quantum state,
$\rho \in {\mathcal
{H}}_{1}\otimes{\mathcal {H}}_{2}\otimes\cdots\otimes{\mathcal
{H}}_{N}$, the corresponding concurrence of (\ref{xxx}) is given by the convex roof:
\begin{eqnarray}\label{def}
\mathcal{C}_{N}(\rho)=\min_{\{p_{i},|\psi_{i}\ra\}}\sum_{i}p_{i}\mathcal{C}_{N}(|\psi_{i}\ra\la\psi_{i}|),
\end{eqnarray}
where the minimum is taken over all the ensemble decomposition of $\rho=\sum_{i}p_{i}|\psi_{i}\ra\la\psi_{i}|$.

The correlation tensors of the generalized Bloch representation of a
quantum state play significant roles in quantum information theory.
In \cite{vicente1,vicente2,hassan,ming}, separable conditions for
both bi- and multi-partite quantum states are introduced by studying
the norm of the correlation tensors. In \cite{hassan1,hassan2}, the
authors present a multipartite entanglement measure for N-qubit and
N-qudit pure states, using the norm of the correlation tensors. In
\cite{vicente3}, the authors have introduced a general framework for
detecting genuine multipartite entanglement and non full
separability in multipartite quantum systems of arbitrary dimensions
based also on the correlation tensors. In \cite{mingbell}, we have
found that the norms of the correlation tensors are closely related
to the maximal violation of a kind of multipartite Bell
inequalities.

In the following, we first reform the concurrence for multipartite pure
states in terms of the norms of the correlation tensors. The correlation tensors
are then used to derive a lower bound of concurrence for mixed multipartite quantum
states. The lower bound also provides a fully separable condition
for multipartite quantum states. We further show that genuine multipartite
entanglement can be detected by the bound. We also investigate the
multipartite tangle. Tight lower and upper bounds are derived.

\section{Lower bound of multipartite concurrence}\label{sec2}

We first consider the concurrence of
multipartite pure states $|\psi\ra \in {\mathcal
{H}}_{1}\otimes{\mathcal {H}}_{2}\otimes\cdots\otimes{\mathcal
{H}}_{N}$ in terms of the generalized Bloch representation of $|\psi\ra\la\psi|$.
Let $\{\lambda_{\alpha_{k}}\}$ be the $SU(d)$ generators.
The generalized Bloch representation for any quantum states $\rho\in {\mathcal {H}}_{1}\otimes{\mathcal
{H}}_{2}\otimes\cdots\otimes{\mathcal {H}}_{N}$ is given by \cite{hassan},
\begin{eqnarray}\label{OS}
\rho=\frac{1}{d^N}\left(\otimes_{j=1}^{N}I_{d}+\cdots
+\sum_{M=1}^N\sum\limits_{\{\mu_{1}\mu_{2}\cdots\mu_{M}\}}\sum\limits_{\alpha_{1}\alpha_{2}\cdots\alpha_{M}}
{\mathcal{T}}_{\alpha_{1}\alpha_{2}\cdots\alpha_{M}}^{\{\mu_{1}\mu_{2}\cdots\mu_{M}\}}\lambda_{\alpha_{1}}
^{\{\mu_{1}\}}\lambda_{\alpha_{2}}^{\{\mu_{2}\}}\cdots\lambda_{\alpha_{M}}^{\{\mu_{M}\}}
\right),
\end{eqnarray}
where $\{\mu_{1}\mu_{2}\cdots\mu_{M}\}$ is a subset of
$\{1,2,\cdots, N\}$,
$\lambda_{\alpha_{k}}^{\{\mu_{k}\}}=I_{d}\otimes
I_{d}\otimes\cdots\otimes \lambda_{\alpha_{k}}\otimes
I_{d}\otimes\cdots\otimes I_{d}$ with $\lambda_{\alpha_{k}}$ appearing
at the $\mu_k$th position and
\begin{eqnarray}\label{dt}
{\mathcal{T}}_{\alpha_{1}\alpha_{2}\cdots\alpha_{M}}
^{\{\mu_{1}\mu_{2}\cdots\mu_{M}\}}=\frac{d^M}{2^{M}}{\rm
Tr}[\rho\lambda_{\alpha_{1}}
^{\{\mu_{1}\}}\lambda_{\alpha_{2}}^{\{\mu_{2}\}}\cdots\lambda_{\alpha_{M}}^{\{\mu_{M}\}}],
\end{eqnarray}
which can be viewed as the entries of the correlation tensors
${\mathcal{T}}^{\{\mu_{1}\mu_{2}\cdots\mu_{M}\}}$.

After computation(detailed processes can be found in the supplementary material), we obtain an alternative representation about the concurrence,
\begin{eqnarray}\label{cfp12}
&&2^{N-2}\mathcal{C}_{N}^2(|\psi\ra\la\psi|)\nonumber\\
&&=[2^N-\frac{(d+1)^N}{d^N}-\frac{1}{d^N}(d^N-1)(d+1)^{N-1}]\nonumber\\
&&\ \ \ \ \ \ +\sum_{l=2}^N\frac{2^l}{d^{N+l}}[(d+1)^{N-1}-(d+1)^{N-l}]\sum_{k_1\cdots
k_l\subset\{1,2,\cdots,N\}}||{\mathcal{T}}^{k_1\cdots k_l}||^2.
\end{eqnarray}

By noticing that a quantum state $\rho$ is fully separable if and only if $\mathcal{C}_{N}(\rho)=0$, one derives a sufficient and necessary condition for the fully separability of multipartite pure states with formula (\ref{cfp12}).
In particular, as the tensors ${\mathcal{T}}^{\{\mu_{1}\mu_{2}\cdots\mu_{M}\}}$ in (\ref{dt}) are
mean values of the observables $\lambda_{\alpha_{1}}^{\{\mu_{1}\}}\lambda_{\alpha_{2}}^{\{\mu_{2}\}}\cdots\lambda_{\alpha_{M}}^{\{\mu_{M}\}}$,
(\ref{cfp12}) also gives an experimental way to measure the concurrence of a pure multipartite state. Since we have eliminated the terms containing
${\mathcal{T}}^{k_1}$, the measurement can be only operated on the norms of ${\mathcal{T}}^{\{\mu_{1}\mu_{2}\cdots\mu_{M}\}}$ with $M\geq2$.
From (\ref{cfp12}) we can now derive the lower bound for multipartite concurrence of any mixed states $\rho$.

{\bf{Theorem 1:}} For any mixed quantum state $\rho \in
{\mathcal {H}}_{1}\otimes{\mathcal
{H}}_{2}\otimes\cdots\otimes{\mathcal {H}}_{N}$, we have
\begin{eqnarray}\label{cfp3}
\mathcal{C}_{N}(\rho)&\geq&
-2^{1-N/2}[-2^N+\frac{(d+1)^N}{d^N}+\frac{1}{d^N}(d^N-1)(d+1)^{N-1}]^{\frac{1}{2}}\nonumber\\
&&+2^{1-N/2}\{\sum_{l=2}^N\frac{2^l}{d^{N+l}}[(d+1)^{N-1}-(d+1)^{N-l}]\sum_{k_1\cdots
k_l\subset\{1,2,\cdots,N\}}||{\mathcal{T}}^{k_1\cdots k_l}||^2\}^{\frac{1}{2}}.
\end{eqnarray}

{\bf{Proof:}} For simplicity we denote
$C=-2^N+\frac{(d+1)^N}{d^N}+\frac{1}{d^N}(d^N-1)(d+1)^{N-1}$, and
$C_{\alpha}$ the coefficient of $||{\mathcal{T}}^{\alpha}||^2$
in (\ref{cfp12}) for $\alpha\in\{k_1k_2, k_1k_2k_3,\cdots,1\cdots
N\}$, which are nonnegative numbers depending only on $N$ and $d$.

Assume that $\rho=\sum_ip_i|\psi_i\ra\la\psi_i|$ is the optimal
decomposition such that (\ref{def}) attains the minimum. We have that
\begin{eqnarray*}
\mathcal{C}_{N}(\rho)&=&\sum_ip_iC_N(|\psi_i\ra)=2^{1-N/2}\sum_ip_i\{-C+\sum_{\alpha}C_{\alpha}||{\mathcal{T}}_i^{\alpha}||^2\}^{\frac{1}{2}}\\
&\geq&2^{1-N/2}[\sum_ip_i\sqrt{\sum_{\alpha}C_{\alpha}||{\mathcal{T}}_i^{\alpha}||^2}-\sqrt{C}]\\
&\geq&2^{1-N/2}[\sqrt{\sum_{\alpha}C_{\alpha}(\sum_ip_i||{\mathcal{T}}_i^{\alpha}||)^2}-\sqrt{C}]\\
&\geq&2^{1-N/2}[\sqrt{\sum_{\alpha}C_{\alpha}||{\mathcal{T}}^{\alpha}||^2}-\sqrt{C}],
\end{eqnarray*}
where we have used the inequalities $\sqrt{a-b}\geq\sqrt{a}-\sqrt{b}$ for $a\geq
b \geq 0$ and $\sum_i\sqrt{\sum_jx^2_{ij}}\geq\sqrt{\sum_j(\sum_ix_{ij})^2}$
for real and nonnegative $x_{ij}$. \hfill \rule{1ex}{1ex}

The lower bound (\ref{cfp3}) can be used to estimate
the concurrence for multipartite quantum states with arbitrary dimension.
It is also a kind of entanglement witness for fully separability. Moreover,
this multipartite concurrence can be employed to detect the genuine multipartite entanglement.
It has been shown that an $N$-partite quantum state $\rho \in{\mathcal
{H}}_{1}\otimes{\mathcal {H}}_{2}\otimes\cdots\otimes{\mathcal
{H}}_{N}$ is genuine multipartite entangled if \cite{GE}
\begin{eqnarray}\label{theorem}
C_{N}(\rho)>
\left\{\begin{array}{l}
\displaystyle 2^{1-\frac{N}{2}}\sqrt{2^{N}-4+\frac{2}{d}-2\sum_{k=1}^{\frac{N-1}{2}}\frac{\binom{N}{k}}{d^k}},~~~~~~~~~~~~~~ \text{for~odd} ~N,\\[6mm]
\displaystyle
2^{1-\frac{N}{2}}\sqrt{2^{N}-4+\frac{2}{d}-2\sum_{k=1}^{\frac{N}{2}-1}\frac{\binom{N}{k}}{d^k}-\frac{\binom{N}{\frac{N}{2}}}{d^{\frac{N}{2}}}},~~~~\text{for~even}
~N,
\end{array}\right.
\end{eqnarray}
where $\binom{N}{k}=N!/(k!(N-k)!)$.

Since the concurrence $\mathcal{C}_{N}(\rho)$ is difficult to compute, our lower bound can be employed to detect
the genuine multipartite entanglement.

As an example, let us consider tripartite case. From (\ref{theorem})
$\rho \in{\mathcal {H}}_{1}\otimes{\mathcal {H}}_{2}\otimes{\mathcal
{H}}_{3}$ is genuinely multipartite entangled if $\mathcal{C}_{3}(\rho)>
\sqrt{2-\frac{2}{d}}$. For a three-qubit GHZ state mixed with noise,
$\rho_{GHZ}=\frac{x}{8}I+(1-x)|GHZ\ra\la GHZ|$, where
$|GHZ\ra=\frac{1}{\sqrt{2}}(|000\ra+|111\ra)$, we have $\mathcal{C}_{3}(\rho_{GHZ})\geq \frac{1}{2}\sqrt{6-25x+\frac{25}{2}x^2}$ by Theorem 1.
Therefore, the lower bound is valid to detect genuinely multipartite entangled for
$x<0.08349$. By $\frac{1}{2}\sqrt{6-25x+\frac{25}{2}x^2}> 0$, one can detect general entanglement(not fully separable) of $\rho_{GHZ}$
for $x<0.27889$. Note that general entanglement will be detected for $x<0.64645$ in \cite{hassan}.

\section{Bounds on multipartite tangle}

We now consider the multipartite tangle that is tightly related with concurrence.
By the squared I-concurrence for bipartite
quantum systems \cite{tangle}, we introduce the multipartite
squared I-concurrence. For a multipartite pure quantum state
$|\psi\ra\in {\mathcal {H}}_{1}\otimes{\mathcal
{H}}_{2}\otimes\cdots\otimes{\mathcal {H}}_{N}$ the multipartite
squared I-concurrence is defined by the square of the multi-concurrence,
\begin{eqnarray}\label{msic}
\tau_N(|\psi\ra\la\psi|)=\mathcal{C}_{N}^2(|\psi\ra\la\psi|)=2^{2-N}[(2^{N}-2)-\sum_{\alpha}Tr\{\rho_{\alpha}^{2}\}],
\end{eqnarray}
where $\alpha$ labels all the different reduced density matrices.  For a
mixed multipartite quantum state,
$\rho=\sum_{i}p_{i}|\psi_{i}\ra\la\psi_{i}| \in {\mathcal
{H}}_{1}\otimes{\mathcal {H}}_{2}\otimes\cdots\otimes{\mathcal
{H}}_{N}$, the corresponding multipartite squared I-concurrence is then given by the convex roof:
\begin{eqnarray}\label{tdef}
\tau_{N}(\rho)=\min_{\{p_{i},|\psi_{i}\ra\}}\sum_{i}p_{i}\tau_{N}(|\psi_{i}\ra\la\psi_{i}|).
\end{eqnarray}
The multipartite squared I-concurrence defined above has the following properties:
(i) $\tau_{N}(\rho)=0$ if and only if $\rho$ is fully separable;
(ii) $\tau_{N}(\rho)$ is invariant under local unitary transformation of $\rho$;
(iii) $\tau_{N}(\rho)\geq \mathcal{C}_{N}^2(\rho)$.
By property (i) above, a multipartite state is not separable if
$\tau_{N}(\rho)>0$. In the following, we present valid lower and upper
bounds for $\tau_{N}(\rho)$.

{\bf{Theorem 2:}} For any mixed quantum state $\rho \in
{\mathcal {H}}_{1}\otimes{\mathcal
{H}}_{2}\otimes\cdots\otimes{\mathcal {H}}_{N}$, we have
\begin{eqnarray}\label{cfp4}
\tau_{N}(\rho)&\geq&
-2^{2-N}[-2^N+\frac{(d+1)^N}{d^N}+\frac{1}{d^N}(d^N-1)(d+1)^{N-1}]\nonumber\\
&&+2^{2-N}\{\sum_{l=2}^N\frac{2^l}{d^{N+l}}[(d+1)^{N-1}-(d+1)^{N-l}]\sum_{k_1\cdots
k_l\subset\{1,2,\cdots,N\}}||{\mathcal{T}}^{k_1\cdots k_l}||^2\};\\
\tau_{N}(\rho)&\leq& 2^{2-N}(2^N-2-\sum_{\alpha} Tr \rho_{\alpha}^2).
\end{eqnarray}

{\bf{Proof:}} We still take the simplified notions $C$ and
$C_{\alpha}$ used in the proof of Theorem 1.
Assume that $\rho=\sum_ip_i|\psi_i\ra\la\psi_i|$ is the optimal
decomposition such that (\ref{tdef}) attains the minimum. We have that
\begin{eqnarray*}
\tau_{N}(\rho)&=&\sum_ip_i\mathcal{C}_N^2(|\psi_i\ra)=2^{2-N}\sum_ip_i\{-C+\sum_{\alpha}C_{\alpha}||{\mathcal{T}}_i^{\alpha}||^2\}\\
&=&2^{2-N}[\sum_{\alpha}C_{\alpha}(\sum_ip_i||{\mathcal{T}}_i^{\alpha}||)^2-C]\\
&\geq&2^{2-N}[\sum_{\alpha}C_{\alpha}||{\mathcal{T}}^{\alpha}||^2-C],
\end{eqnarray*}
where we have used the triangle inequality for the Hilbert-Schmidt norm.

On the other hand, by the definition of $\tau_{N}(\rho)$, we have
\begin{eqnarray*}
\tau_{N}(\rho)&\leq&\sum_ip_i\tau_N(|\psi_i\ra)=2^{2-N}(2^N-2-\sum_{\alpha,i}p_iTr(\rho_{\alpha}^i)^2)\\
&\leq& 2^{2-N}[2^N-2-\sum_{\alpha} Tr (\sum_ip_i\rho_{\alpha}^i)^2]\\
&=& 2^{2-N}(2^N-2-\sum_{\alpha} Tr \rho_{\alpha}^2),
\end{eqnarray*}
which gives the upper bound. \hfill \rule{1ex}{1ex}

From the proof of Theorem 2, one has that for
pure states the lower and upper bounds are exact. Thus the
lower and upper bounds (\ref{cfp4}) for $\tau_N(\rho)$ are tight.

{\it Remark:} Our bounds are given by the norms of the
correlation tensors. As the Hilbert-Schmidt norm is invariant under
local unitary transformation, the bounds give
experimentally feasible way in identify both non-separability and genuine multipartite
entanglement. Further more, as has been discussed in
\cite{vicente3, hassan}, partial knowledge of the correlation
tensors may also allow us to detect entanglement and estimate the degree of entanglement.

\section{Conclusions and Discussions}\label{sec5}

It is a basic and fundamental question in quantum entanglement
theory to compute the concurrence for multipartite quantum
systems. Since the concurrence is defined by taking the optimization
over all the ensemble decompositions of a mixed quantum states, it
is formidable to derive an analytical formula. We have derived an analytical and
experimentally feasible formula for multipartite concurrence
of any multipartite pure quantum states by using generalized
Bloch representation of density matrices.
We have then obtained a lower bound of concurrence for any
mixed multipartite quantum states. Genuine multipartite entanglement can be detected by
using this bound. We have also investigated the multipartite tangle. Tight
lower and upper bounds are obtained. Although the bounds are in general not tighter than that in \cite{jpali},
they are experimentally feasible as the elements in the correlation tensors are just the mean values of the hermitian $SU(d)$ generators.
The approach used in this manuscript
can also be implemented to investigate the k-separability of multipartite
quantum systems. Future research on
the construction of genuine multipartite entanglement criteria
in terms of the lower bound of multipartite
squared I-concurrence and the k norm would be also interesting. The correlation tensors, based on which the results
are derived in this paper, would be used further to investigate nonlocality\cite{mingbell} or Measurement-Induced Non locality\cite{mil}.

\bigskip
\noindent{\bf Acknowledgments}\, \,This work is supported by the
NSFC 11105226, 11275131; the Fundamental Research Funds for the
Central Universities No. 15CX08011A, 15CX05062A, 27R1410016A, 16CX02049A;
Qingdao applied basic research program
No. 15-9-1-103-jch, and a project sponsored by SRF for
ROCS, SEM.

\section*{Supplemental material for " Bounds on Multipartite Concurrence and Tangle"}

\subsection{Equivalent representation of multipartite concurrence for pure states}

To rewritten $\mathcal{C}_{N}(|\psi\ra\la\psi|)$, one can represent both ${\rm Tr}\rho^2$
and ${\rm Tr}\rho^2_{\alpha}$ with respect to the correlation tensors.
By using the orthogonality of the $SU(d)$ generators we have
\begin{eqnarray}\label{tr1}
{\rm Tr}\rho^2=\frac{1}{d^{2N}}(d^N+
\sum_{M=1}^N2^Md^{N-M}\sum_{k_1\cdots
k_M\subset{\{1,2,\cdots,N\}}}||{\mathcal{T}}^{k_1\cdots k_M}||^2)
\end{eqnarray}
and
\begin{eqnarray}\label{redm}
{\rm Tr}\rho^2_{k_1\cdots
k_M}=\frac{1}{d^{2M}}(d^M+
\sum_{r=1}^M2^rd^{M-r}\sum_{k_{j_1}\cdots
k_{j_r}\subset{\{k_{j_1}\cdots
k_{j_M}\}}}||{\mathcal{T}}^{k_{j_1}\cdots
k_{j_r}}||^2)
\end{eqnarray}
for any $1\leq M\leq N-1$, where $\rho_{k_1\cdots k_M}$ is the $M$-partite reduced density
matrix supporting on ${\mathcal {H}}_{k_1}\otimes{\mathcal
{H}}_{k_2}\otimes\cdots\otimes{\mathcal {H}}_{k_M}$.

For any $1\leq M\leq N-1$, we get
\begin{eqnarray*}
&&\sum_{k_1,k_2,\cdots,k_M}({\rm Tr}\rho^2-{\rm Tr}\rho^2_{k_1\cdots
k_M})\\
&&=\sum_{l=0}^M(\binom{N}{M}\frac{2^ld^{N-l}}{d^{2N}}-\binom{N-l}{M-l}\frac{2^ld^{M-l}}{d^{2M}})\sum_{k_1k_2\cdots k_l\subset\{1,2,\cdots,N\}}||{\mathcal{T}}^{k_1k_2\cdots k_l}||^2\\
&&+\sum_{l=1}^{N-M}\binom{N}{M}\frac{2^{M+l}d^{N-(M+l)}}{d^{2N}}\sum_{k_1k_2\cdots k_{M+l}\subset\{1,2,\cdots,N\}}||{\mathcal{T}}^{k_1k_2\cdots k_{M+l}}||^2.
\end{eqnarray*}

Notice that for pure state, we have ${\rm Tr}\rho^2=1$. Thus the concurrence for pure state can be equivalently represented by
\begin{eqnarray}\label{eqcp}
\mathcal{C}_{N}(|\psi\ra\la\psi|)=2^{1-\frac{N}{2}}\sqrt{\sum_{\alpha}(Tr \rho^2-Tr\{\rho_{\alpha}^{2}\})}.
\end{eqnarray}
By substituting the representation for $\sum_{k_1,k_2,\cdots,k_M}({\rm Tr}\rho^2-{\rm Tr}\rho^2_{k_1\cdots
k_M})$ above into (\ref{eqcp}), for pure states $|\psi\ra$ we have
\begin{eqnarray}\label{cfp}
&&2^{N-2}\mathcal{C}_{N}^2(|\psi\ra\la\psi|)\nonumber\\
&&=\sum_{M=1}^{N-1}\sum_{k_1,k_2,\cdots,k_M}({\rm Tr}\rho^2-{\rm Tr}\rho^2_{k_1\cdots
k_M})\nonumber\\
&&=\frac{1}{d^N}[-(d+1)^N+d^N+2^N-1]\nonumber
\\&&\ \ \ \ \ \
+\sum_{l=1}^N\frac{2^l}{d^{N+l}}[-(d+1)^{N-l}+2^N-1]\sum_{k_1k_2\cdots k_l\subset\{1,2,\cdots,N\}}||{\mathcal{T}}^{k_1k_2\cdots k_l}||^2
\end{eqnarray}

Since ${\rm Tr}\rho^2=1$ for any pure state $\rho=|\psi\ra\la\psi|$, from (\ref{tr1}) we have
\begin{eqnarray}\label{tr2}
\sum_{k_1\in\{1,2,\cdots N\}}||{\mathcal{T}}^{k_1}||^2=
\frac{d^{2N}-d^N}{2d^{N-1}}
-\sum_{l=2}^N\frac{2^ld^{N-l}}{2d^{N-1}}\sum_{k_1\cdots
k_l\subset\{1,2,\cdots,N\}}||{\mathcal{T}}^{k_1\cdots k_l}||^2.
\end{eqnarray}
Substituting (\ref{tr2}) into (\ref{cfp}), we obtain an alternative relation about the the concurrence,
i.e. equation (\ref{cfp12}).

\end{document}